%% file: main.tex
\documentclass[twocolumn,reprint,amsmath,amssymb,footinbib,superscriptaddress,floatfix,aps,longbibliography,prl]{revtex4-2}
\usepackage[sort&compress]{natbib}
\usepackage{colortbl} 
\usepackage{graphicx}
\usepackage{dcolumn}
\usepackage{bm}
\usepackage{physics}
\usepackage{dsfont}
\usepackage{hyperref}

\usepackage[dvipsnames]{xcolor}

\usepackage{blindtext}

\usepackage{mwe}

\usepackage{ulem}

\hypersetup{
  colorlinks   = true, 
  urlcolor     = ForestGreen,
  linkcolor    = ForestGreen,
  citecolor   = ForestGreen 
}

\begin{document}

\title{Zero-field dipolar decoupling of color center ensembles via universal qutrit control}

\author{Antonio Verdú}
\email{a.verdugomariz(at)um.es}
\affiliation{Departamento de Física, Universidad de Murcia, 30071 Murcia, Spain}
\affiliation{Monodon, Navantia, S.A, S.M.E., 28005 Madrid, Spain}

\author{Ana Teresa Gea-Caballero}
\affiliation{Departamento de Física, Universidad de Murcia, 30071 Murcia, Spain}

\author{Santiago Oviedo-Casado}
\affiliation{Universidad Politécnica de Cartagena member of European University of Technology EUT+, Área de Física Aplicada, Departamento de Física Aplicada y Tecnología Naval, 30202 Cartagena, Spain}
\affiliation{Escuela Superior de Ingeniería y Tecnología, Universidad Internacional de La Rioja, 26006 Logroño, La Rioja, Spain}

\author{Fedor Jelezko}
\email{fedor.jelezko(at)uni-ulm.de}
\affiliation{Institute for Quantum Optics, Ulm University, Albert-Einstein-Allee 11, 89081 Ulm, Germany}
\affiliation{Center for Integrated Quantum Science and Technology (IQST), 89081 Ulm, Germany}

\author{Javier Prior}
\email{javier.prior(at)um.es}
\affiliation{Departamento de Física, Universidad de Murcia, 30071 Murcia, Spain}

\begin{abstract}

Dipolar interactions are a major source of decoherence in dense ensembles of color centers in diamond. Current protocols demand using bias magnetic fields detrimental in many scenarios. We present ZENITH (Zero-field Ensemble Neutralization via Interleaved Trilevel Handling), a pulsed sequence to cancel dipolar interactions among V-degenerate spin-1 systems. We reveal excellent coherence survival, compatibility with existing sensing sequences, and improved DC detection, advancing a general framework to control interacting degenerate multilevel systems, of broad interest in quantum technologies.

\end{abstract}

\maketitle

\begin{figure}
    \centering
    \includegraphics[width= \columnwidth]{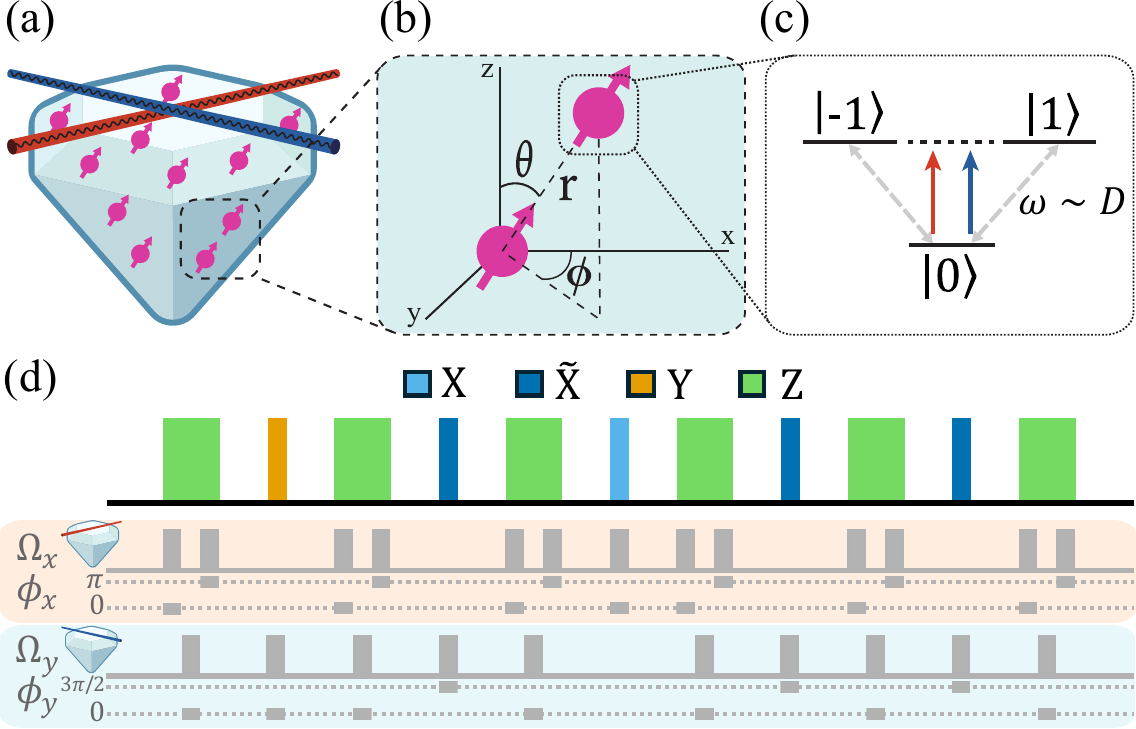}
    \caption{\textbf{ZENITH schematics} (a)  Experimental setup representation for arbitrary control of an ensemble of color centers. Two perpendicular microwave sources, orthogonal to the NV centers symmetry axis, enable universal control of the NV center ground state qutrit manifold. 
    (b) Two-body dipolar interaction among electron spins, described here in relative spherical coordinates $\{r,\theta,\phi\}$.
    (c) A driving that matches the V-degenerate qutrit's zero-field splitting $\omega \sim D$ enables double-state operators that create different superpositions of the states $\ket{1}$ and $\ket{-1}$. 
    (d) ZENITH sequence of $\pi/2$ rotations on different axes for dipolar coupling removal.  Microwave rotations are implemented through pulses with tailored amplitudes and phases on each control channel, as indicated by the amplitude-phase scheme in the diagram.}
    \label{fig:figure1}
\end{figure}

Solid-state quantum systems such as color centers in diamond underpin emerging quantum technologies. The naturally accessible quantum properties, ease of controllability, and versatility of applications has meant an accelerated development of various types of implanted defects in the diamond crystal structure.
Possessing a natural V-like spin-1 triplet (qutrit) ground state manifold with degenerate $\ket{\pm 1}$ states shown in Fig.\,\ref{fig:figure1}(c), the NV center
is one of the most prominent quantum magnetometers 
\cite{Rugar2013,Muller2014,Cappellaro2015,Bucher2023}, offering a long coherence time at ambient conditions \cite{Lukin2006,Jelezko2009,Walsworth2010}. 

Usually, a bias magnetic field aligned with the NV center axis is applied to lift the $\ket{\pm 1}$ degeneracy via Zeeman splitting \cite{Cappellaro2017}, thus attaining an excellent qubit approximation in which to perform high-fidelity quantum operations. On such qubits, on-site noise mitigation protocols greatly increase coherence survival \cite{Viola1999,Sarma2008,Ryan2010,Souza2011,Suter2016,Genov2017}, unlocking the NV center as a precise quantum sensor \cite{Dolde2011,Neumann2013,Cappellaro2015}, among other applications \cite{Childress2013}. Furthermore, suppression of interaction-induced noise --such as the dipolar coupling presented in Fig.\,\ref{fig:figure1}(b)-- in qubit arrays is also possible through homonuclear control sequences, including WAHUHA \cite{Bar-Gill2018, Balasubramanian2019}, which render ensembles as collections of isolated spins and potentially increase sensitivity \cite{Barry2020}.

Although decoupling sequences where originally designed for qubits, generalization to multi-level systems exist \cite{OKeeffe2019,Zhou2024}. Nevertheless, these sequences target systems with well separated energy levels, failing when used in NV center ensembles with no bias magnetic field applied. Yet these
magnetic fields brings important caveats, such as masking weak interactions \cite{Sutter2012}, or causing structural changes in analyte molecules \cite{Thiel2016,Pelliccione2016,Jenkins2019}. Operating at zero bias field allows the sensor to access the J-coupling rich spectra of molecules \cite{Blanchard2013,Barskiy2025,Omar2026}, is crucial for environment quantum control \cite{Bauch2018,Ajoy2019}, and permits improving vector-field magnetometry \cite{Childress2025}, eliminating systematic errors and simplifying the experimental setup \cite{Schwartz2019}. 
Nonetheless, although controlling degenerate states in isolated qutrits is possible through specifically designed pulses \cite{Sekiguchi2016,Saijo2018,Zheng2019,Cerrillo2021,Lenz2021,Vetter2022,Li2024,Jiang2024}, a dipolar decoupling sequence of pulses capable of working at zero bias field with strongly interacting qutrit ensembles is lacking. Our article fulfills this gap.

In this article, we present ZENITH (Zero-field Ensemble Neutralization via Interleaved Trilevel Handling), a pulsed sequence to decouple dipolar interactions between color centers in dense ensembles at zero bias magnetic field.  
We demonstrate the existence of a sequence of eleven pulses, shown in Fig.\,\ref{fig:figure1}(d), that cancels the dipolar interaction, and use numerically exact techniques \cite{PhysRevLett.107.070601}, to reveal extended coherence survival in qutrit ensembles. Moreover, we show smooth integrability of ZENITH with a Ramsey-like sensing sequence, leading to enhanced DC sensitivity.
The generality of our derivation makes it broadly applicable to nearly degenerate three-level systems across diverse platforms beyond NV centers.

\begin{figure*}[t]
    \centering
    \includegraphics[width=\linewidth]{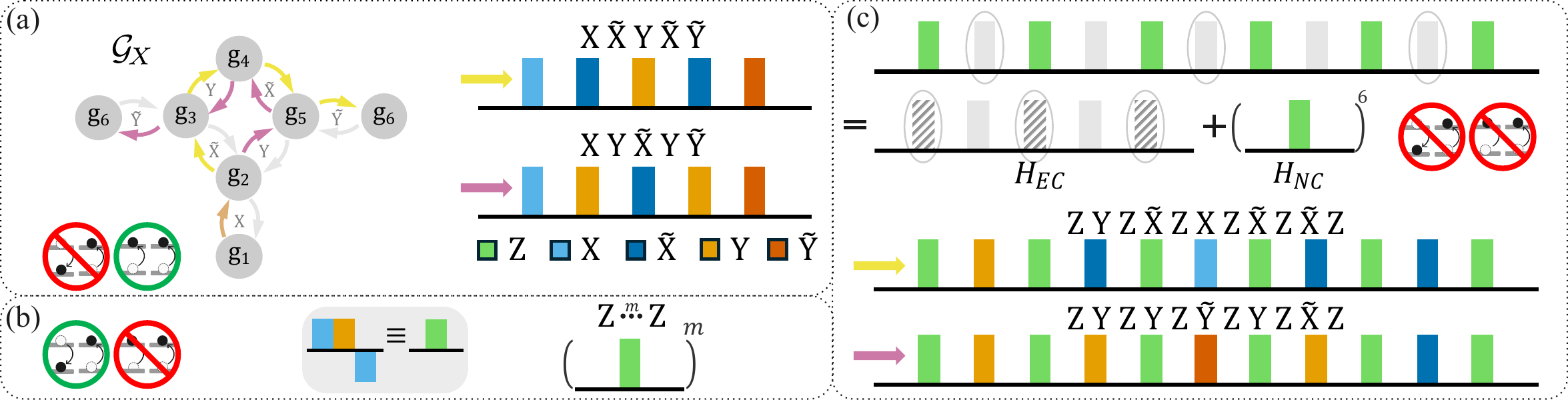}
    \caption{\textbf{Pictorial derivation of ZENITH dipolar decoupling sequence}
    (a) Construction of the sequence that decouples the energy-conserving dipolar Hamiltonian $H_{EC}$ in Eq.\,\eqref{eq:hamiltonian_sf}. Minimal graph representing the connections between the different pulsed-transformed versions of $H_{EC}$, denoted as $g_i \in \mathcal{G}_X$ (balls) through the set of $\pi/2$ rotations $V=\{X,Y,\Tilde{X},\Tilde{Y}\} $ (arrows), starting with a free-evolution time $g_1=\mathds{1}$. All minimal solutions are achieved by two equally long paths (yellow and pink arrows), both starting from the same pulse (orange arrow).
    (b) Decoupling the non-energy conserving dipolar Hamiltonian $H_{NC}$  requires from $Z$ pulses, obtained through the compositions $ABA^{-1}$ and $B^{-1}AB$,
    where the pair $(A,B)$ corresponds to either $(X,Y)$ or $(\Tilde{X},\Tilde{Y})$.
    (c) Full dipolar decoupling sequence integration. (Top): When $Z$ rotations are integrated in the $H_{EC}$ minimal decoupling sequence in (a) operators in odd positions alter as: $X,\Tilde{X} \leftrightarrow Y,\Tilde{Y}$, while $H_{NC}$ is canceled just by a train of balanced $Z$ pulses. (Bottom): representation of the two distinct pulses sequences for complete dipolar decoupling. 
    All $\pi/2$ pulses are represented equally, although they can be realized as direct pulses or implemented through composite pulses.
    }
    \label{fig:figure2}
\end{figure*}

\textit{Full control of V-qutrits.---}
The Hamiltonian describing an ensemble of V-symmetrical qutrits with dipole-dipole interaction is
\begin{eqnarray}
    H = \sum_i \left[D(S_z^{(i)})^2 + H_c^{(i)}(t) \right] + \sum_{i\ne j}  H_{dd}^{(ij)},
\end{eqnarray}
where $D$ is the zero-field splitting that separates the $\ket{0}$ state from the degenerate $\ket{\pm 1}$ states, and $S_z$ is the spin-1 operator along the quantization axis. Our aim is to find the minimal sequence of pulse operators that cancels the dipolar interaction $H_{dd}$. Working in the rotating frame with respect to $ D \left[(S_z^{(i)})^2+(S_z^{(j)})^2\right]$, and ignoring the fast rotating terms, the dipolar interaction Hamiltonian splits into $H_{dd}'^{(ij)}=H_{EC}^{(ij)}+H_{NC}^{(ij)}$, involving energy-conserving (EC) and non energy-conserving (NC) transitions, which are, respectively,
\begin{align}
    H_{EC}^{(ij)}&= \frac{J_0(1-3\cos^2\theta)}{4r^3}\left(4S_z^{(i)}S_z^{(j)}-S_x^{(i)}S_x^{(j)} \nonumber \right.
    \\& \left. -S_y^{(i)}S_y^{(j)}-\Tilde{S}_x^{(i)}\Tilde{S}_x^{(j)}-\Tilde{S}_y^{(i)}\Tilde{S}_y^{(j)}\right),  \label{eq:hamiltonian_sf}\\
    H_{NC}^{(ij)}&= \frac{-3J_0\sin^2\theta e^{-i2 \phi}}{2r^3} ( \ketbra{10}{0 \,\text{-}1} + \ketbra{01}{\text{-}10})+ \text{H.c.}
    \label{eq:hamiltonian_df}
\end{align}
Here, $J_0 = (2\pi) \,52 \, \text{MHz}\cdot \text{nm}^3$ is the interaction strength, and $\{r,\theta,\phi\}$ the relative distance, and polar and azimuthal angles between each interacting spin pair [see Fig.\,\ref{fig:figure1}(b)]. Eq.\,\eqref{eq:hamiltonian_sf} includes single flip-flop or phase flip terms, and is described by the use of spin operators and the anticommutators $\Tilde{S}_{x,y}=\{S_{x,y}, S_z\}$, while Eq.~\eqref{eq:hamiltonian_df} includes double flip-up/down transitions. The latter term, while effectively suppressed under high-field conditions due to fast dynamics \cite{Zhou2024,Choi_2017}, becomes a distinctive feature in the zero- or low-bias magnetic-field regime and must be explicitly addressed.

To achieve arbitrary control over the ensemble, we consider external microwave radiation delivered through two orthogonal cables located perpendicularly to the NV's symmetry axis $\hat{z}$, as shown in Fig.\,\ref{fig:figure1}(a) \cite{London2014}. The electromagnetic fields couple to the qutrits through $H_c^{(i)}(t) = \gamma \mathbf{B}(t)\cdot\mathbf{S}^{(i)}$, with $\gamma$ the electron's gyromagnetic ratio, and $B_{x,y}=\Omega_{x,y}(t)\cos(\omega t + \phi_{x,y})$ the transverse components of the driving fields.
Both microwave signals have frequencies chosen to match the NV zero-field splitting $\omega \sim D$  [Fig.\,\ref{fig:figure1}(c)] \cite{Cerrillo2021, Lpez-Garca2025}, and the amplitude $\Omega_{x,y}(t)$ and phase $\phi_{x,y}$ are axis-specific. Working in the rotating frame with respect to $ D S_z^2$, ignoring the fast rotating terms through a rotating wave approximation, assuming that $\Omega(t) \ll D$, we get the control
\begin{equation}\label{eq:control}
\begin{split}
       H_c'(t) &= \frac{\gamma}{2}\left\{\Omega_x(t)\left[\cos(\phi_x)S_x+\sin(\phi_x)\tilde{S}_y\right] \right.\\&\left. +\Omega_y(t)\left[\cos(\phi_y)S_y-\sin(\phi_y)\tilde{S}_x\right]\right\},
\end{split}
\end{equation}
which permits universal manipulation of a qutrit through the appropriate choice of amplitudes and phases for each microwave source, as shown in the Supplemental Material (SM) \cite{supp_information}.

\textit{Zero-field dipolar decoupling.---} To design a pulsed control sequence that removes the dipolar coupling, we employ Average Hamiltonian Theory (AHT), such that the zeroth-order average Hamiltonian vanishes, thereby suppressing the leading-order contribution to the effective dynamics \cite{Haeberlen1968}. We address the two terms of the dipolar interaction separately. First, we design a base sequence that cancels the energy-conserving part $H_{EC}$ in Eq.\,\eqref{eq:hamiltonian_sf}. We then upgrade it to additionally suppress the non–energy-conserving terms $H_{NC}$ in Eq.\,\eqref{eq:hamiltonian_df}.  

To engineer the base protocol, we use the fact that the four $\pi/2$ rotations generated by the double-transition operators $\mathcal{S}=\{S_x,S_y,\Tilde{S}_x,\Tilde{S}_y\}$ swap the set $\mathcal{S}'=\mathcal{S}\cup\{ S_z, \Tilde{S}_z \}$ in a closed manner \cite{supp_information}, with the anticommutator $\Tilde{S}_z=\{S_x,S_y\}$.
These unitary operators correspond to linear microwave driving on each of the orthogonal directions to the NV axis. 
Owing to the two-body nature of the interaction, and the simultaneous application of the operations to all qutrits, only the absolute value of the resulting transformation is relevant. Therefore, equivalent solutions arise from choosing either $\pm \pi/2$ rotations. 
We define the set of
rotations as $V=\{X,Y,\Tilde{X},\Tilde{Y}\} = \bigotimes_{j} e^{- i \frac{\pi}{2}\mathcal{S}^{(j)}}$.

For a sequence of $l-1$ equally spaced pulses, the zeroth-order average Hamiltonian of $H_{EC}$ reads $\Bar{H}^{(0)}=\frac{\tau}{T}\sum_i  T_i H_{EC} T_i^\dagger$, where $T_i$ represents each toggling frame transformation and $\tau$ is the duration of each interval. Then, $T= l \tau$ is the total protocol time. The transformations $T_i$ are obtained as chronologically ordered products of the applied physical pulses $P_j$, i.e., $T_i = P_1\cdot P_2 \cdot P_3 \cdot ... \cdot P_i$. The decoupling procedure can be described within an abstract group-theoretical framework \cite{Viola1999,Viola2003}: The transformations $T_i$ are analogous to the group elements $g_i$ of a finite group $G$, generated by the elements $v_\alpha \in V$. The action of $G$ is then equivalent to projecting onto the $G$-invariant component of $H_{EC}$, defined as $\Pi_G(H_{EC})=\frac{1}{|G|}\sum_{g_i\in G} g_i H_{EC}g_i^\dagger$ , where $|G|$ is the order of the group. Minimal decoupling sequences are then subsets $\mathcal{G}_\nu \subset G$ such that $\Pi_{\mathcal{G}_\nu}(H_{EC})=0$.
We identify such subsets through a search based on the construction of the Cayley graph, $\Gamma(G,V)$, built by joining the elements of $G$ (vertices) through the elements of $V$ (directed edges), i.e., the pathways or pulses that allow to arrive at any given frame considered. 

Choosing to start with a free evolution, corresponding to $g_1=\mathds{1}$, 
we find four minimal subsets $\mathcal{G}_{v_0}$ of order six that decouple $H_{EC}$, with $v_0\in V$ denoting the initial pulse. For clarity, here we restrict the analysis to the subset $\mathcal{G}_X$ and discuss the generalization in Sec. IV of the SM \cite{supp_information}. In Fig.\,\ref{fig:figure2}(a) we display the minimal subgraph of the full $\Gamma(G,V)$ containing the decoupling group $\mathcal{G}_X$. Decoupling consists of traversing all the elements $g_i$ of $\mathcal{G}_{X}$, with $i\in[1,6]$. From the graph structure, two distinct paths can be identified, each allowing traversing the graph's vertices through direct jumps, i.e, through single pulses. Hence, we obtain a family of five-pulses sequences that decouples $H_{EC}$.

For $H_{NC}$, we utilize the properties of global $\pi/2$ rotations about the $z$ axis ($Z$). These rotations are accessible through the control Hamiltonian in Eq.\,\eqref{eq:control} by means of composite pulses, as both sets $\{S_x, S_y\}$ and $\{\Tilde{S}_x, \Tilde{S}_y\}$ close the spin-1 algebra with the operator $S_z$. The $Z$ rotations generate a $\mathbb{Z}_2$ symmetry under which $H_{EC}$ is even, whereas $H_{NC}$ is odd, satisfying $ZH_{EC}Z^\dagger=H_{EC}$ and $ZH_{NC}Z^\dagger=-H_{NC}$. 
As a result, a sequence of intermediate $Z$ pulses divides the total evolution into multiple free-evolution intervals $\tau$, in which the non-energy-conserving contribution alternates between $H_{NC}$ and $-H_{NC}$. If the pulse timings are chosen such that the total time spent in both signs is equal, the contributions compensate in the average Hamiltonian. Therefore, repetition of $m$ balanced $(\tau\text{-}Z\text{-}\tau)$ blocks suppresses the non-energy-conserving dynamics while preserving the energy-conserving evolution (Fig.\,\ref{fig:figure2}(b)). 

The different transformation properties of $H_{EC}$ and $H_{NC}$ under $Z$ rotations also provide the route for full decoupling. Intermediate $Z$ pulses isolate either contribution, allowing additional control pulses to be inserted between balanced $Z$ blocks without reintroducing non-energy-conserving contributions. The train of $Z$ rotations modifies the effective toggling-frame transformations of the new pulses at the odd positions of the sequence \cite{supp_information}, as represented in Fig.\,\ref{fig:figure2}(c). Therefore, the full decoupling sequence is obtained by adding five pulses whose transformed action reproduces the $H_{EC}$ decoupling solution in Fig.\,\ref{fig:figure2}(a). The ZENITH protocol is then described as a family of eleven-pulse sequences [Fig.\,\ref{fig:figure2}(c)], with the freedom to choose the starting pulse and an arbitrary phase $\varphi = \{ -\frac{\pi}{2}, \frac{\pi}{2}\}$ for each of the rotations.

\begin{figure}
    \centering
    \includegraphics[width=\linewidth]{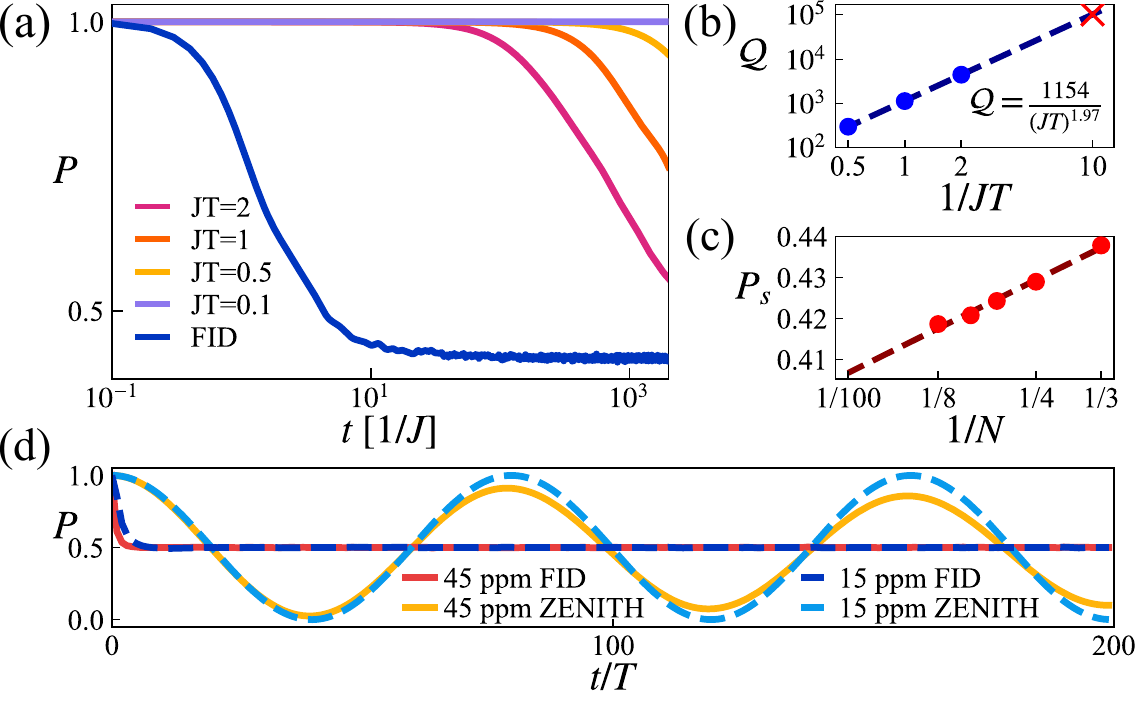}
    \caption{\textbf{ZENITH performance}
    (a) Numerical time evolution of $P$, computed as the average of the survival probabilities of the states $\ket{\uparrow}$,$\,\,\ket{\downarrow}$, and $\ket{+}$,
    on an ensemble of eight randomly distributed dipolar interacting qutrits under FID (solid blue line) and with the stroboscopic implementation of ZENITH for values of $JT = 2,1,0.5, 0.1$, respectively. For the faster decaying uncontrolled evolution and $JT=2$ curves we average 10 cluster profiles to reach convergence, while for the rest a single cluster profile is enough, as no significant variations occur for the considered duration.
    (b) Protection factor as a function of $JT$. The factor for the protocol with $JT = 0.1$ is estimated to be $\mathcal{Q}\sim106526$ (red mark) as a result of the fit.
    (c) Finite-size effect study. For the evolution without ZENITH, representation of the saturation population $P_s$ between $Jt=(492,500)$ for $N=3,4,5,6,8$. To reduce the effect of fluctuations, we take the mean between $(600,600,500,200,70)$ different cluster configurations.
    (d) Simulated response of the average of 20 cluster profiles of eight NVs at two different NV concentrations (15 and 45 ppm) to a DC signal of coupling strength $\mu T= 0.1$, considering FID and a ZENITH protocol with optimized initial state.}
    \label{fig:figure3}
\end{figure}

\textit{Numerical verification.---}
We numerically characterize the performance of ZENITH relative to an uncontrolled free-induction decay (FID) by measuring the average survival probability $P$ of three distinct qutrit coherent initial states: $\ket{\uparrow}=\frac{1}{\sqrt{2}}(\ket{0}+\ket{+1})$,$\,\,\ket{\downarrow}=\frac{1}{\sqrt{2}}(\ket{0}+\ket{-1})$, and $\ket{+}=\frac{1}{\sqrt{2}}(\ket{+1}+\ket{-1})$.
With the experimental performance in mind, we simulate a symmetrized protocol by appending the reversed sequence to the end, a standard technique for suppressing odd-order terms in the average Hamiltonian \cite{mansfield1971symmetrized,Brinkmann2016}.
Being $U_T$ the unitary of duration $T$ describing the system's time evolution, in Fig.\,\ref{fig:figure3}(a), we evaluate the stroboscopic dynamics $(U_T)^n$, $n\in \mathbb{Z}^+$, with and without ZENITH, and display the time evolution of $P$ for 
different values of $JT$ ($0.1, 0.5, 1, 2$), where $J = \frac{J_0}{r_0^3}$ and $r_0$ is the minimum distance between particles in the clusters.
We realize unitary time evolution using the Time-Dependent Variational Principle \cite{PhysRevLett.107.070601} on sets of eight NV centers following a homogeneous 3D distribution and average the results for different numbers of cluster profiles until we obtain convergence.

The sequence demonstrates excellent performance, showing trivial evolution for a long time following state initialization. 
We quantify this performance through a protection factor $\mathcal{Q} = t_z/t_f$ where $t_z$ and $t_f$ are the times at which the average survival probability with ZENITH and FID, respectively, decay to $95\%$ of its initial value. Across the considered interaction range, $\mathcal{Q} \sim 290-10^5$.
The parameter exhibits a power-law dependence $\mathcal{Q} \propto (JT)^{-\alpha}$ (see Fig.\,\ref{fig:figure3}(b)), with scaling exponent $\alpha=1.97 \pm 0.02$, which can be used to estimate the expected protection factor for a given concentration or to determine the optimal sequence duration.
For example, considering a representative strongly interacting ensemble, with an NV center concentration of 45 ppm, or dipolar interaction strength of $J\approx (2\pi) \,420\, \text{kHz}$ \cite{Kucsko2018, Choi_2017}, we get a protection factor $\mathcal{Q}\approx 337$ using experimentally accessible parameters of pulse separation $\tau = 20$ ns and pulse duration $t_p = 5$ ns \cite{London2014}, thereby gaining more than two orders of magnitude in average coherence time. The same experimental parameters in a typical 15 ppm ensemble \cite{Zhou2020} lead to $\mathcal{Q}\approx 2900$.

The accessible system size in our simulations is limited to clusters of, at most, eight particles, owing to the computational cost of simulating all-to-all interactions \cite{Hastings_2007, Schollw_ck_2011}.
To assess finite-size effects, Fig.\,\ref{fig:figure3}(c) shows the long-time asymptotic population average $P_s$ as a function of particle number $N$. The observed approximate $1/N$ scaling agrees with standard finite-size corrections in mean-field theory \cite{Liberti2010}. At $N=8$, the deviation from the infinite-size limit is only $\sim10^{-2}$, supporting the robustness of the results in Fig.\,\ref{fig:figure3}(a).

\textit{DC magnetometry.---}
A central requirement of any proposed dipolar decoupling solution is compatibility with quantum sensing protocols. To assess its achievement with ZENITH, we simulate the response of an NV center ensemble to a longitudinally coupled static (DC) signal, described as $H_s = \mu S_z$, with $\mu$ the signal strength. Under $H_s$, any given sequence of the ZENITH family leads to a complex, three-frequency response \cite{supp_information}. However, a Ramsey-like experiment in which the initial pulse creates an equally weighted superposition state of the two eigenvectors with the largest energy difference, yields a signal oscillating at a single frequency dependent on the signal strength and the maximal energy separation, which for ZENITH is fixed at $\Delta \lambda = \frac{1}{2\sqrt{3}}(1+\sqrt{3}) \approx 0.789$ \cite{supp_information}. Then, detection of the target signal requires estimating the frequency of oscillation, just as in conventional Ramsey, with the advantage of having a significantly larger coherence time.

Figure\,\ref{fig:figure3}(d) displays stroboscopic evolution on a cluster of eight NVs targeting a DC signal of strength $\mu=0.1/T$, both for FID and with ZENITH initiated on the optimal measurement state, considering two different experimentally relevant NV ensemble concentrations. While for FID dipolar coupling induces a fast decay of the signal, thereby preventing frequency estimation, the large coherence time availed by ZENITH yields clean oscillation from which the original DC signal can be reconstructed. Note that, as we show in Sec. VI of the SM \cite{supp_information}, the optimal initial state can be achieved with a single pulse, and therefore does not introduce additional errors, nor it reduces the experiment time compared to a FID experiment. 

\textit{Discussions and outlook.---}
ZENITH derivation is valid in the impulsive limit, when the driving frequency is strong enough to mask other contributions during the pulse application time. Away from this limit, several techniques can be implemented to make the sequence robust against sources of extra dynamics during the pulses, to account for the finite duration of the pulses, or to remove higher-order terms in the AHT. Strategies such as using composite pulses \cite{Kabytayev2014}, Pulse-cycle Decoupling \cite{Burum1979,Tyler2023,Zhou2023}, or Eulerian Dynamical Decoupling \cite{Viola2003}, can improve the performance of the sequence, at the price of extending the sequence with several new pulses. The framework employed in this work simplifies the derivation of Eulerian decoupling, as this method involves completing an Eulerian cycle through the complete graph $\Gamma(G,V)$, i.e., a closed path where each edge is visited exactly once before returning to the starting point. Complementary techniques, such as group factorization \cite{Read2025}, could help reduce the complexity of the optimized sequence.

Additional noise sources not accounted for in the derivation, such as strain, stray electromagnetic fields, or magnetic impurities like the P1 centers common in high-density NV ensembles \cite{Levine2018}, can limit coherence survival beyond what ZENITH offers. In these scenarios, full decoupling of the NV center's environment is possible through standard techniques that remove single-site noise effects. Using the dipolar Hamiltonian invariance under $\pi$ pulses, dynamical decoupling processes can be seamlessly integrated in ZENITH. 
Overall, many-body interactions represent a fundamental limitation to zero-field operations with dense ensembles of color centers in diamonds. We solve the problem using a universal set of operations readily accessible with tested experimental setups. Using NV center ensembles with no bias field brings opportunities for biosensing or molecular spectroscopy, as well as applications beyond quantum sensing, making of ZENITH a crucial addition to the quantum technologies toolbox.

\begin{acknowledgments}
\textit{Acknowledgments.---} A.V. and A.T.G.C. contributed equally to this work. We thank Alejandro Martínez Méndez for helpful discussions. A.T.G.C. acknowledges funding from the University of Murcia through the Predoctoral Contracts Program of the Research Promotion Plan (Plan de Fomento de la Investigación). This work was supported by the QuantERA II program (Mf-QDS) and QuantERA III program (AQuSeND) that have received funding from the European Union’s Horizon 2020 research and innovation program under Grant Agreement No. 101017733 and from the Agencia Estatal de Investigación, with project codes PCI2022-132915, PCI2024-153474 and by QNAVIUM Project SCPP2400C011413XV0 funded by MICIU/AEI/10.13039/501100011033 and by the European Union NextGenerationEU/PRTR. And, by the European Union (Quantum Flagship project ASPECTS, Grant Agreement No. 101080167). F.J. is supported by the German Federal Ministry of Research, Technology and Space (BMFTR) via future cluster QSENS ( 03ZU2110DB )  EXTRASENS (13N16935)    DIAQNOS (  13N16463), quNV2.0 (  13N16707),   Deutsche Forschungsgemeinschaft (DFG) via projects   387073854, 445243414, ,  491245864, 546850640, 560722984,  and joint DFG/JST ASPIRE program via project  554644981, European Union's HORIZON Europe program via projects CQuENS (No. 101135359),   and FLORIN (No. 101086142), European Research Council (ERC) via Synergy grant HyperQ (No. 856432), and Carl-Zeiss-Stiftung via Research Center QPhoton and project Ultrasensvir.
\end{acknowledgments}

 \bibliography{references}
\nocite{Bertlmann2008}

\clearpage

\include{supplemental}

\end{document}

%% file: supplemental.tex
\appendix
\counterwithout{equation}{section}
\counterwithout{figure}{section}
\counterwithout{table}{section}

\setcounter{equation}{0}
\setcounter{figure}{0}
\setcounter{table}{0}
\setcounter{page}{1}

\renewcommand{\theequation}{S\arabic{equation}}
\renewcommand{\thefigure}{S\arabic{figure}}
\renewcommand{\thetable}{S\arabic{table}}

\renewcommand{\bibnumfmt}[1]{[S#1]}
\renewcommand{\citenumfont}[1]{S#1}

\widetext
\clearpage
\begin{center}
\textbf{\large Supplemental Materials for ``Zero-field decoupling of color center ensembles via universal qutrit control"}
\end{center}

\section{Section I. Gell-Mann matrices and useful operators}

Working with NV centers in the zero- and low-field regimes generally requires retaining the full three-level ground-state manifold, rather than reducing the dynamics to an effective two-level subspace. The system must therefore be treated as a qutrit, whose general traceless Hermitian operators are spanned by the generators of the $\mathfrak{su}(3)$ Lie algebra. A standard  basis for this algebra, together with the identitity $\mathds{1}_{3\cross3}$, is given by the eight Gell-Mann matrices [\hyperlink{suppref:bertlmann2008}{S1}], defined below as
\begin{eqnarray}
  &  \lambda_1 =
\begin{pmatrix}
0 & 1 & 0 \\
1 & 0 & 0 \\
0 & 0 & 0
\end{pmatrix}, 
\lambda_2 =
\begin{pmatrix}
0 & -i & 0 \\
i & 0 & 0 \\
0 & 0 & 0
\end{pmatrix}, 
\lambda_3 =
\begin{pmatrix}
1 & 0 & 0 \\
0 & -1 & 0 \\
0 & 0 & 0
\end{pmatrix}, 
\lambda_4 =
\begin{pmatrix}
0 & 0 & 1 \\
0 & 0 & 0 \\
1 & 0 & 0
\end{pmatrix}, &\\
&\lambda_5 =
\begin{pmatrix}
0 & 0 & -i \\
0 & 0 & 0 \\
i & 0 & 0
\end{pmatrix},
\lambda_6 =
\begin{pmatrix}
0 & 0 & 0 \\
0 & 0 & 1 \\
0 & 1 & 0
\end{pmatrix}, 
\lambda_7 =
\begin{pmatrix}
0 & 0 & 0 \\
0 & 0 & -i \\
0 & i & 0
\end{pmatrix}, 
\lambda_8 =
\frac{1}{\sqrt{3}}
\begin{pmatrix}
1 & 0 & 0 \\
0 & 1 & 0 \\
0 & 0 & -2
\end{pmatrix}.&
\end{eqnarray}
For convenience, we work with an alternative set of matrices, namely the spin-1 operators and their anticommutators
\begin{eqnarray}
    &S_x = (\lambda_6 + \lambda_1)/\sqrt{2} , \hspace{1cm}
    S_y = (\lambda_2 + \lambda_7)/\sqrt{2}, \hspace{1cm}
    S_z = \frac{1}{2}(\lambda_3 + \sqrt{3}\lambda_8),&\\
    &\Tilde{S}_x = ( \lambda_1 -\lambda_6)/\sqrt{2}= \{S_x,S_z\} ,\hspace{0.5cm}
    \Tilde{S}_y = ( \lambda_2 -\lambda_7)/\sqrt{2}= \{S_y,S_z\}, \hspace{0.5cm}
    \Tilde{S}_z=\lambda_5 = \{S_x,S_y\}.&\nonumber
    \label{eq:operators}
\end{eqnarray}
The design of the protocol is based on $\pi/2$ rotations over the axes described by the set $\mathcal{S}=\{S_x,S_y,\Tilde{S}_x,\Tilde{S}_y\}$, defined as $V=\{X,Y,\Tilde{X},\Tilde{Y}\} =  e^{- i \frac{\pi}{2}\mathcal{S}}$. The spin-1 operators and their anticommutators are transformed in a closed manner under such rotations, following the relations presented in Table\,\ref{tab:rotaciones-pi2}.

\begin{table}[h]
\centering
\begin{tabular}{||c|c |c| c| c| c| c||}
\hline
              & $\mathbf{S_x}$ & $\mathbf{S_y}$ & $\mathbf{S_z}$ & $\mathbf{\tilde{S}_x}$ & $\mathbf{\tilde{S}_y}$ & $\mathbf{\tilde{S}_z}$ \\ \hline

$\mathbf{X}$         & $S_x$ & $S_z$ & $-S_y$ & $\tilde{S}_z$ & $-\tilde{S}_y$ & $-\tilde{S}_x$ \\ 
\hline

$\mathbf{Y}$         & $-S_z$ & $S_y$ & $S_x$ & $-\tilde{S}_x$ & $-\tilde{S}_z$ & $\tilde{S}_y$ \\ 
\hline

$\mathbf{{\tilde X}}$ & $-\tilde{S}_z$ & $-S_y$ & $-\tilde{S}_y$ & $\tilde{S}_x$ & $S_z$ & $S_x$ \\
\hline

$\mathbf{{\tilde Y}}$ & $-S_x$ & $\tilde{S}_z$ & $\tilde{S}_x$ & $-S_z$ & $\tilde{S}_y$ & $-S_y$  \\ \hline 
\end{tabular}
\caption{Transformations under $\pi/2$ rotations in $V$ of spin-1 operators and their anticommutators}
\label{tab:rotaciones-pi2}
\end{table}

\section{Section II. Derivation of the dipolar Hamiltonian}
The general form of the two-body dipole-dipole interaction Hamiltonian is
\begin{equation}
    H_{dd}^{(ij)}= -\frac{J_0}{r^3}\Big[3 (\vec{S}^{(i)}\cdot \hat{n}) (\Vec{S}^{(j)}\cdot \hat{n})-\Vec{S}^{(i)} \cdot \Vec{S}^{(j)}\Big],
    \label{eq:Hdd-initial}
\end{equation}
with $\hat{n}$ the unitary vector connecting both spins. The expression can be expanded into 
\begin{align}
H_{dd}^{(ij)}
= &-\frac{J_0}{r^3}\Big[ 
(3n_x^2-1)S_x^{(i)}S_x^{(j)}
+ (3n_y^2-1)S_y^{(i)}S_y^{(j)}
+ (3n_z^2-1)S_z^{(i)}S_z^{(j)}  \notag \\
& +3n_xn_y(S_x^{(i)}S_y^{(j)}+S_y^{(i)}S_x^{(j)})
+ 3n_xn_z(S_x^{(i)}S_z^{(j)}+S_z^{(i)}S_x^{(j)})
+ 3n_yn_z(S_y^{(i)}S_z^{(j)}+S_z^{(i)}S_y^{(j)})
\Big].
\end{align}

We move to a rotating frame with respect to the zero field splitting $ D (S_z^{2\,(i)}+S_z^{2\,(j)})$ [\hyperlink{suppref:choi2017}{S2}]. As the control consists of global pulses simultaneously applied to each electron spin, we can treat the operator on each site individually and then perform the tensor product. In this frame, the operators are transform through the relations
\begin{eqnarray}
    &S_x^{\prime} =\frac{1}{\sqrt{2}} \Big[ e^{iDt} (\ketbra{1}{0}+ \ketbra{-1}{0}) + e^{-iDt} (\ketbra{0}{1}+ \ketbra{0}{-1})  \Big]& \\
    &S_y^{\prime} =\frac{i}{\sqrt{2}} \Big[ e^{iDt} (-\ketbra{1}{0}+ \ketbra{-1}{0}) + e^{-iDt} (\ketbra{0}{1}- \ketbra{0}{-1})  \Big]& 
\end{eqnarray}
First thing to notice is that the term $S_z$ remains invariant under this transformation, and, therefore, the term $S_z S_z$ will remain unchanged. Also, from the previous expressions, and considering that $D$ is much larger than the rest of the energy scales, all the terms involving tensor products of $S_z$ with both $S_x$ and $S_y$ will oscillate fast, making them negligible under the rotating wave approximation. 
The different tensor products combinations involving $S_x$ and $S_y$ include both time-independent terms and terms oscillating with frequency $2D$. Using the rotating-wave approximation, we neglect the latter, leading to 

\begin{eqnarray}
    &S_x^{\prime} \otimes S_x^{\prime} = \frac{1}{2} \Big[ \ketbra{10}{01} + \ketbra{10}{0-1} + \ketbra{-10}{01} + \ketbra{-10}{0-1} \Big] + \text{H.c.} ,&\\
    &S_y^{\prime} \otimes S_y^{\prime}= \frac{1}{2} \Big[ -\ketbra{10}{01} + \ketbra{10}{0-1} + \ketbra{-10}{01} - \ketbra{-10}{0-1} \Big] + \text{H.c.} ,&\\
    &S_x^{\prime} \otimes S_y^{\prime} = \frac{i}{2} \Big[ \ketbra{10}{01} - \ketbra{10}{0-1} + \ketbra{-10}{01} - \ketbra{-10}{0-1} \Big] + \text{H.c.} ,&\\
    &S_y^{\prime} \otimes S_x^{\prime} = \frac{i}{2} \Big[ -\ketbra{10}{01} - \ketbra{10}{0-1} + \ketbra{-10}{01} + \ketbra{-10}{0-1} \Big]  + \text{H.c.} .& 
\end{eqnarray}

Using polar coordinates $(n_x,n_y,n_z) = (\sin\theta\cos\phi, \sin\theta \sin\phi, \cos\theta)$ and simplifying the expressions, we can rewrite $H^{(ij)}_{dd}$ as
\begin{align}
     H^{\prime(ij)}_{dd}= \ \frac{J}{r^3} &\Big[ \big(1-3\cos^2\theta\big) S_z^{\prime(i)}S_z^{(j)}    \nonumber\\
     & -\frac{3}{2}\sin^2\theta \big(e^{2i\phi} \ketbra{0-1}{10}+ e^{-2i\phi} \ketbra{01}{-10}\big) -\frac{1}{2} (1-3\cos^2\theta) \big( \ketbra{01}{10} + \ketbra{0-1}{-10} \big) \Big] + \text{H.c.},
     \label{eq:Hdd-rf-rwa}
\end{align}

For notational simplicity, we drop the primes and indices and write ${H}^{\prime (ij)}_{dd}$ simply as ${H}_{dd}$. We express Eq.\eqref{eq:Hdd-rf-rwa} as a sum of terms with different physical character, $H_{dd}=H_{EC}+H_{NC}$, where $H_{EC}$ and $H_{NC}$ denote the energy-conserving (EC) and non-energy-conserving (NC) contributions, respectively. The former contains only single flip-flop and phase-flip terms, whereas the latter includes double flip-up/down transitions. We write these terms as

\begin{eqnarray}
    &H_{EC}= \frac{J_0(1-3\cos^2\theta)}{4r^3}\left(4S_z^{(i)}S_z^{(j)}-S_x^{(i)}S_x^{(j)} 
      -S_y^{(i)}S_y^{(j)}-\Tilde{S}_x^{(i)}\Tilde{S}_x^{(j)}-\Tilde{S}_y^{(i)}\Tilde{S}_y^{(j)}\right), \nonumber   \label{eq_supp:hamiltonian_sf}&\\
    &H_{NC}^{(ij)}= \frac{-3J_0\sin^2\theta e^{-i2 \phi}}{2r^3} ( \ketbra{10}{0 \,\text{-}1} + \ketbra{01}{\text{-}10})+ \text{H.c.}&
    \label{eq_supp:hamiltonian_df}
\end{eqnarray}

\section{Section III. Zero-field control}

To describe the achievable arbitrary control of the qutrit sensor in the zero-field regime, we start from a Hamiltonian describing the coupling between an NV center and two sources of electromagnetic radiation orthogonal to its symmetry ($z$) axis
\begin{eqnarray}
     H_c(t) = \gamma \mathbf{B}(t)\cdot\mathbf{S}=\gamma(\Omega_x(t)\cos(\omega t+\phi_x)S_x+\Omega_y(t)\cos(\omega t+\phi_y)S_y),
\end{eqnarray}
where $\Omega_{x,y}(t)$ are the time-dependent amplitudes of each of the sources, $\phi_{x,y}$ are the corresponding phases and $\gamma$ is the electron's gyromagnetic ratio. In the rotating frame of the zero-field splitting $DS_z^2$ [\hyperlink{suppref:cerrillo2021}{S3},\hyperlink{suppref:vetter2022}{S4}] and applying rotating wave approximation with the choice of driving frequency $\omega=D$, the Hamiltonian acquires the form 
\begin{eqnarray}
    H'_c(t) =& \frac{\gamma}{2}\Big[ \Omega_x(t)( e^{-i\phi_x} \ketbra{+}{0}) -
    i \Omega_y(t)e^{-i\phi_y}\ketbra{-}{0}\Big]+\text{H.c.},
\end{eqnarray}
written in the $\{\ket{+},\ket{-}\}$ basis, where $\ket{\pm}=\frac{1}{\sqrt{2}}(\ket{+1}\pm\ket{-1})$. It is also possible to write the expression more compactly in terms of appropriate operators, in particular
\begin{eqnarray}
    H_c'(t) = \frac{\gamma}{2}\left\{\Omega_x(t)\left[\cos(\phi_x)S_x+\sin(\phi_x)\tilde{S}_y\right] \right.\left. +\Omega_y(t)\left[\cos(\phi_y)S_y-\sin(\phi_y)\tilde{S}_x\right]\right\}.
\end{eqnarray}

Once we have the control Hamiltonian, we investigate its effect on the system's populations. Although several variables are present in the Hamiltonian, we can easily calculate the exponential due to a closed form of the powers of the control Hamiltonian. In particular, the condition $H_c^3=(\Omega_x^2+\Omega_y^2)H_c$ is fulfilled, allowing us to write the exponential $U(t)=e^{-i t H_c}$ as

\begin{eqnarray}
    U(t)=\mathds{1}-\frac{i\sin(\Omega t)}{\Omega}H_c+\frac{\cos(\Omega t)-1}{\Omega^2}H_c^2,
\end{eqnarray}
with $\Omega^2=\Omega_x^2+\Omega_y^2$. If we calculate the value of $H_c^2$:
\begin{eqnarray}
    H_c^2=\Omega^2 \ketbra{0}{0}+2\Omega_x^2\ketbra{+}{+}+2\Omega_y^2\ketbra{-}{-}+2ie^{-i(\phi_x-\phi_y)}\Omega_x\Omega_y \ketbra{+}{-}+\text{H.c.}
\end{eqnarray}

The final expression for the evolution operator under two orthogonal microwave driving is then
\begin{eqnarray}
    U(t)=&&\cos(\Omega t)\ketbra{0}{0}+\ketbra{+}{+}+\ketbra{-}{-}+
    \frac{\cos(\Omega t)-1}{\Omega^2}(\Omega_1^2\ketbra{+}{+}+\Omega_2^2\ketbra{-}{-})\\
    &&-i\frac{\sin(\Omega t)}{2 \Omega}(\Omega_xe^{-i\phi_x}\ketbra{+}{0}+\Omega_ye^{-i\phi_y}\ketbra{-}{0})+
    \frac{2i\Omega_1\Omega_2}{\Omega^2}(\cos(\Omega t)-1)e^{-i(\phi_x-\phi_y)}\ketbra{+}{-}+\text{H.c.}\nonumber
\end{eqnarray}

From the previous expression, we can show direct access to an arbitrary combination in the double-quantum subspace $\ket{\chi}=a\ket{1}+b\ket{-1}$ with a single pulse after initialization onto the $\ket{0}$ state. This is helpful, for example,  in initial state preparation. We seek a pulse of duration $T$ such that there is no population leakage to $\ket{0}$. Consequently, we need to avoid the $\ketbra{0}{0}$ components in $U(T)$. To accomplish this, we need the condition $\cos(\Omega T)=0$ to be satisfied. Therefore, if we let the system evolve for $T=\frac{\pi}{2\Omega}$, there is no remaining population on $\ket{0}$. Applying the evolution operator for time $T$ to state $\ket{0}$, leads to
\begin{eqnarray}
    U_c(T)\ket{0}=\frac{-i}{2\Omega}(e^{-i\phi_x}\Omega_x\ket{+}+e^{-i\phi_y}\Omega_y\ket{-}),
\end{eqnarray}
describing an arbitrary state in the double-quantum subspace.

\section{Section IV. Family of solutions of the energy-conserving terms}

In the main text we analyze $\mathcal{G}_X$, one of the four minimal subsets that produce decoupling of $H_{EC}$. Here we present the other three subgroups $\mathcal{G}_{Y},\mathcal{G}_{\tilde{X}},\mathcal{G}_{\tilde{Y}}$ and the relations between them. The existence of this family of solutions comes from the symmetric form of $H_{EC}$ with respect to the generators of the rotations. To get more intuition, we write the Hamiltonian in the following form
\begin{eqnarray}
    H_{EC}=J \, \text{diag}(-1,-1,4,-1,-1,0) \mathbf{S}^{(i)}(\mathbf{S}^{(j)})^T
\end{eqnarray}
where $J=\frac{J_0(1-3\cos^2\theta)}{4r^3}$ and $\mathbf{S}=(S_x,S_y,S_z,\Tilde{S}_x,\Tilde{S}_y,\Tilde{S}_z)$. All the operators involved appear tensored with themselves. Furthermore, we see that all the elements from the set $\mathcal{S}$ present the same weight in $H_{EC}$. This concept is the basis that leads to a family of solutions.

\begin{figure}
    \centering
    \includegraphics[width=1\linewidth]{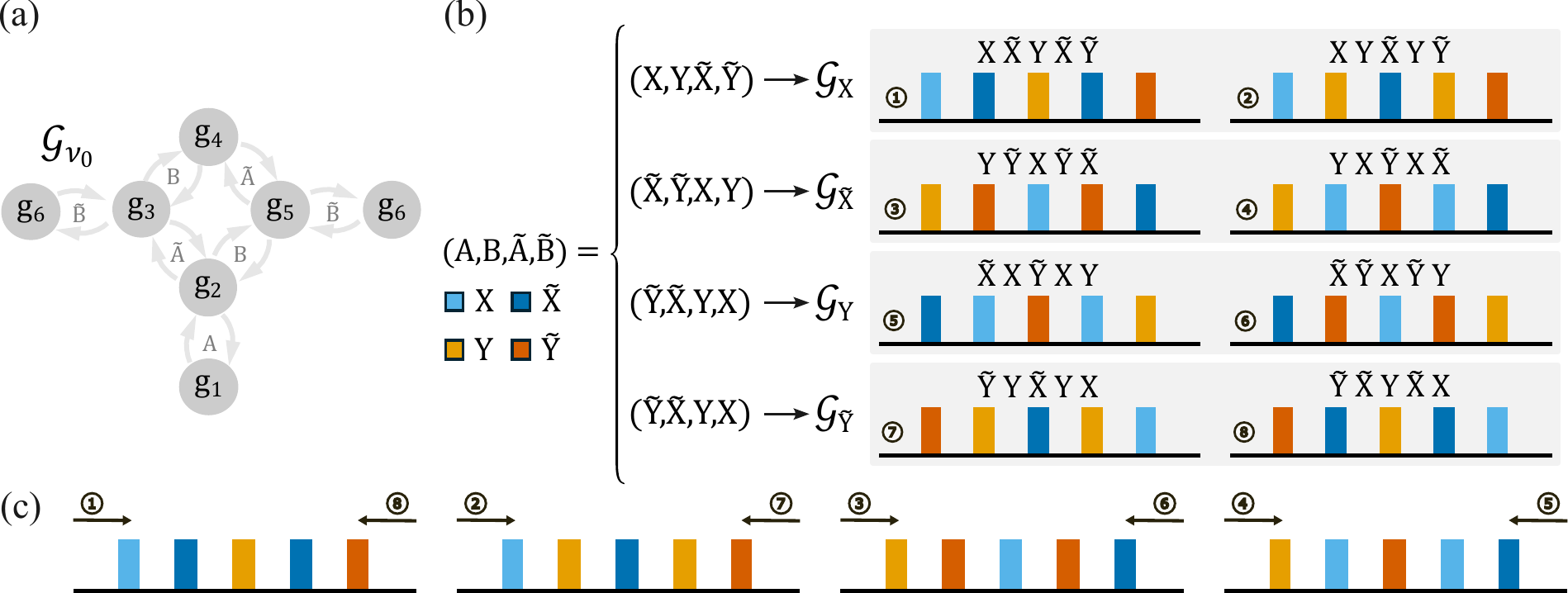}
    \caption{ \textbf{Family of decoupling sequences of $H_{EC}$}
    (a) General minimal subgraph that produces decoupling. The structure is valid for any initial pulse $A \in V$, providing two different decoupling routes in each case.
    (b) Full set of eight decoupling pulse sequences. 
    (c) Due to the symmetry of solutions, the full set of solutions can be grouped in four different sequences that can be traversed either forwards or backwards. }
    \label{fig_supp:cayley_generalized}
\end{figure}

The same effect is reflected in the Cayley graph of the group. We can see that the structure represented in the main text is valid for the different initial pulses, with their corresponding decoupling groups. In Fig.\,\ref{fig_supp:cayley_generalized}(a) we represent the generalization of the minimal subgraph that produces decoupling, which is valid for the four different choices of initial pulse $A=(X,Y,\Tilde{X},\Tilde{Y})$. The same pattern of pulses is repeated in all of the subgraphs, which we represent by means of the dummy indices $(A,B,\Tilde{A},\Tilde{B})$, which can present the values $(X,Y,\Tilde{X},\Tilde{Y})$, $(Y,X,\Tilde{Y},\Tilde{X})$, $(\Tilde{X},\Tilde{Y},X,Y)$ or $(\Tilde{Y},\Tilde{X},Y,X)$. In a pulsed sequence form, we represent in Fig.\,\ref{fig_supp:cayley_generalized}(b) all the set of eight solutions to the averaging of $H_{EC}$. Furthermore, we observe that the solutions can be grouped in pairs that can be traversed either forward or backward, as the doublets $(\mathcal{G}_X,\mathcal{G}_{\Tilde{Y}})$ and $(\mathcal{G}_Y,\mathcal{G}_{\Tilde{X}})$ present symmetric solutions (Fig.\,\ref{fig_supp:cayley_generalized}(c)). This leads to a generalization of all the possibilities in four pulsed sequences that can be traversed in both directions. 

\section{Section V. Derivation of the full sequence}

By means of the Cayley graphs, we derived the sequences that average out the part of the dipolar interaction that consist of energy-conserving terms. Additionally, we know that intermediate $\pi/2$ pulse in the $z$ direction ($Z$ pulses) can remove the part that contains non-energy-conserving terms while keeping the energy-conserving part invariant. Using that property, we show how the addition of intermediate $Z$ pulses in between of the original sequence produces full decoupling, subjected to a modification in the odd pulses of the sequence.  Let $H_{dd}$ be the total dipolar Hamiltonian, $P_j$ the j-th pulse of a sequence and $Z_i$ the i-th pulse in the $z$ direction. To prove the validity of the sequence, we start calculating the toggling-frame Hamiltonians one by one.
\begin{eqnarray}
    &H_0 = H_{dd}; \,
    H_1 = Z_1 H_{dd} Z_1^\dagger;\,
    H_2=Z_1 P_1 H_{dd} P_1^\dagger Z_1^\dagger.&
\end{eqnarray}
In the last expression, we can add an identity $Z_1^\dagger Z_1$
\begin{eqnarray}
    H_2=Z_1 P_1 Z_1^\dagger Z_1 H_{dd} Z_1^\dagger Z_1 P_1^\dagger Z_1^\dagger= P_1'Z_1 H_{dd} Z_1^\dagger (P_1')^\dagger,
\end{eqnarray}
where $P_i'$ is defined as a $\pi/2$ rotation over the $z$ axis acting on the pulse $P_i$. The next term will be
\begin{eqnarray}
    H_3 = P_1'Z_1Z_2 H_{dd}Z_2^\dagger Z_1^\dagger (P_1')^\dagger=P_1'H_{dd} (P_1')^\dagger,
\end{eqnarray}
where we note that $Z_1 Z_2$ leaves $H_{dd}$ unchanged, independently of the phases of the individual rotations due to its global rotation character. The following contributions are 
\begin{eqnarray}
    &H_4 = P_1'P_2H_{dd}P_2^\dagger (P_1')^\dagger, \,
    H_5 = P_1'P_2 Z_3 H_{dd}Z_3^\dagger P_2^\dagger (P_1')^\dagger,&
\end{eqnarray}
recovering back the structure of $H_0$ and $H_1$. This pattern will be repeated throughout the whole sequence. Therefore, our new partial Hamiltonians will all include transformed operators in the odd positions (1,3,5 in our case). Furthermore, using that $R_z H_{dd} R_z^\dagger = R_z^\dagger H_{dd} R_z$ allows to split the Hamiltonian into two groups of terms that be treated separately. Defining $H = H_{dd}+ R_z H_{dd} R_z^\dagger$, we can calculate the zeroth order average Hamiltonian as
\begin{eqnarray}
    &\Bar{H}^{(0)}= \sum_{i=0}^{11} H_i = J[H] = H+ (P_1') H (P_1')+(P_1')(P_2) H (P_2)^\dagger(P_1')^\dagger+
   (P_1')(P_2) (P_3') H (P_3')^\dagger (P_2)^\dagger(P_1')^\dagger+& \nonumber\\
    &(P_1')(P_2) (P_3')(P_4) H (P_4)^\dagger(P_3')^\dagger (P_2)^\dagger(P_1')^\dagger+
    (P_1')(P_2) (P_3')(P_4) (P_5') H (P_5')^\dagger (P_4)^\dagger(P_3')^\dagger (P_2)^\dagger(P_1')^\dagger,&
\end{eqnarray}
where we defined the effect of the pulses as a functional $J[\bullet]$ acting on some Hamiltonian.
The previous expression has the form of a sequence consisting of five-pulses acting on the newly defined $H$. We consider now that the set of rotations $\{P_1',P_2,P_3',P_4,P_5'\}$ fulfills the decoupling condition for $H_{Ec}$. Splitting the dipolar interaction into its two constituent parts, we can show full dipolar decoupling
\begin{eqnarray}
    \Bar{H}^{(0)}=J[H] = J[H_{dd}+R_zH_{dd}R_z^\dagger] =J[H_{EC}] +J[H_{NC}]+J[R_z H_{EC}R_z^\dagger]+J[R_z H_{NC}R_z^\dagger].
\end{eqnarray}
Using that $R_z H_{EC}R_z^\dagger = H_{EC}$ and  $R_z H_{NC}R_z^\dagger = -H_{NC}$, we get
\begin{eqnarray}
    \Bar{H}^{(0)}= 2 J[H_{EC}]+J[H_{NC}+R_z H_{NC} R_z^\dagger] = 0,
\end{eqnarray}
showing that the zeroth-order dipolar interaction is fully removed.

\section{Section VI. Signal response and optimal initial-state preparation}

We consider the response of the NV sensor to a longitudinally coupled DC signal,
described by the Hamiltonian
\begin{equation}
    H_s = \mu S_z .
\end{equation}
Under the application of any sequence belonging to the ZENITH family, this signal is
mapped onto an effective Hamiltonian of the form
\begin{equation}
    H_{\mathrm{eff}}
    =\frac{\mu}{6}(
    \eta_x S_x
    +\eta_y S_y
    +\eta_z S_z
    +\mu_{xz}\{S_x,S_z\}
    +\mu_{yz}\{S_y,S_z\}
    +\mu_{xy}\{S_x,S_y\}).
\end{equation}
where the coefficients $\eta_\alpha,\mu_{\beta \gamma}=\pm 1$ depend on the particular sequence. In particular, due to the initial free-evolution under $H_s$, $eta_z$ always presents the value $+1$. Thus, the
sequence does not completely remove the signal, but rather redistributes it over
different operators. This redistribution reduces the effective signal amplitude.
We quantify this reduction through the spectral range of \(H_{\mathrm{eff}}\), namely the
difference between its largest and smallest eigenvalues.

To show that this reduction is universal for the whole family of sequences, we first
remove the overall prefactor \(\mu/6\) and define
\begin{equation}
    h= S_z
    \eta_x S_x
    +\eta_y S_y
    +\mu_{xz}\{S_x,S_z\}
    +\mu_{yz}\{S_y,S_z\}
    +\mu_{xy}\{S_x,S_y\}.
\end{equation}
The eigenvalues of \(h\) are obtained from its characteristic equation,
\begin{equation}
    \lambda^3
    -\mathrm{Tr}(h)\lambda^2
    +\frac{1}{2}
    \left[
    \mathrm{Tr}(h)^2-\mathrm{Tr}(h^2)
    \right]\lambda
    -\frac{1}{6}
    \left[
    \mathrm{Tr}(h)^3
    -3\mathrm{Tr}(h)\mathrm{Tr}(h^2)
    +2\mathrm{Tr}(h^3)
    \right]
    =0 .
\end{equation}
In the present case, \(h\) is traceless and satisfies \(\mathrm{Tr}(h^2)=12\), so this
reduces to
\begin{equation}
    \lambda^3 - 6\lambda - \frac{1}{3}\mathrm{Tr}(h^3)=0 .
\end{equation}
The only remaining sequence-dependent quantity is therefore \(\mathrm{Tr}(h^3)\). For
the Hamiltonians generated by the ZENITH family, one finds
\begin{equation}
    \mathrm{Tr}(h^3)
    =
    -6\left[
    \mu_{yz}\left(\eta_y+\mu_{xz}\mu_{xy}\right)
    +
    \eta_x\left(\mu_{xz}-\eta_y \mu_{xy}\right)
    \right].
\end{equation}
Since all coefficients take values \(\pm1\), this expression can only take the two
values \(\mathrm{Tr}(h^3)=\pm12\). Indeed, writing the expression inside brackets as
\begin{equation}
    S=A(B+CD)+E(C-BD),
\end{equation}
with \(A,B,C,D,E=\pm1\), either \(CD=B\) or \(CD=-B\). In the first case,
\(S=2A\), while in the second case \(S=2E\). Hence \(S=\pm2\), which implies
\(\mathrm{Tr}(h^3)=\pm12\). The characteristic equation is therefore always one of the two cubics
\begin{equation}
    \lambda^3-6\lambda\pm4=0 .
\end{equation}
For this equation (depressed cubic), there exist an analytical trigonometric formula for the three solutions
\begin{eqnarray}
   \lambda_k=2\sqrt{2}\cos(\frac{1}{3}\arccos(\pm \frac{1}{\sqrt{2}})+\frac{2\pi k}{3}),\, k =0,1,2.
\end{eqnarray}
For \(\lambda^3-6\lambda+4=0\), the eigenvalues are
\begin{equation}
    \lambda = 2, \qquad -1+\sqrt{3}, \qquad -1-\sqrt{3},
\end{equation}
whereas for \(\lambda^3-6\lambda-4=0\), they are
\begin{equation}
    \lambda = -2, \qquad 1+\sqrt{3}, \qquad 1-\sqrt{3}.
\end{equation}
In both cases, the difference between the largest and smallest eigenvalues is
\begin{equation}
    \Delta \lambda = 3+\sqrt{3}.
\end{equation}
Reintroducing the prefactor \(\mu/6\), the effective signal frequency associated with
the largest spectral separation is
\begin{equation}
    \Delta \lambda_{\mathrm{eff}}
    =
    \frac{\mu}{6}(3+\sqrt{3})
    =
    \mu \frac{1+\sqrt{3}}{2\sqrt{3}} .
\end{equation}
Thus, all ZENITH sequences produce the same signal-reduction factor,
\begin{equation}
    \mathcal{R}
    =
    \frac{\Delta \lambda_{\mathrm{eff}}}{\mu}
    =
    \frac{1+\sqrt{3}}{2\sqrt{3}} .
\end{equation}

We now determine the initial state that optimizes the sensing response. Let
\(\{\ket{v_i}\}\) be the eigenbasis of \(H_{\mathrm{eff}}\), with corresponding
eigenvalues \(E_i\), and consider a general initial state
\begin{equation}
    \ket{\psi}
    =
    a\ket{v_1}
    +
    b\ket{v_2}
    +
    c\ket{v_3}.
\end{equation}
The survival probability under evolution with \(H_{\mathrm{eff}}\) is
\begin{align}
    P(t)
    &=
    |\braket{\psi}{\psi(t)}|^2  \nonumber \\
    &=
    |a|^4+|b|^4+|c|^4
    +2|a|^2|b|^2\cos(\omega_{12}t)
    +2|a|^2|c|^2\cos(\omega_{13}t)
    +2|b|^2|c|^2\cos(\omega_{23}t),
\end{align}
where \(\omega_{ij}=E_i-E_j\). Equivalently,
\begin{equation}
    P(t)
    =
    1
    -
    4\left[
    |a|^2|b|^2
    \sin^2\left(\frac{\omega_{12}t}{2}\right)
    +
    |a|^2|c|^2
    \sin^2\left(\frac{\omega_{13}t}{2}\right)
    +
    |b|^2|c|^2
    \sin^2\left(\frac{\omega_{23}t}{2}\right)
    \right].
\end{equation}
A state mixture of the three eigenvectors generally contains three different
oscillation frequencies. To obtain a single-frequency response, the initial state must
therefore be restricted to a superposition of only two eigenstates. The optimal choice
is the pair of eigenstates associated with the largest spectral separation,
\(\ket{v_{\max}}\) and \(\ket{v_{\min}}\). The oscillation amplitude is maximized when
both states enter with equal weight, giving
\begin{equation}
    \ket{\psi_{\mathrm{opt}}}
    =
    \frac{1}{\sqrt{2}}
    \left(
    \ket{v_{\max}}
    +
    e^{i\phi}\ket{v_{\min}}
    \right),
\end{equation}
where the relative phase \(\phi\) can be chosen according to the desired readout
quadrature.

It remains to show that these optimal states can be prepared with the available
control. For this purpose, it is useful to write the effective Hamiltonian in matrix form
\begin{equation}
    H_{\mathrm{eff}}
    =
    \begin{pmatrix}
        \eta_z & \alpha & -i\mu_{xy}\\
        \alpha^*  & 0 & \beta\\
        i\mu_{xy} & \beta^* & -\eta_z
    \end{pmatrix},
\end{equation}
with
\begin{equation}
    \alpha
    =
    \frac{1}{\sqrt{2}}
    \left[
    \eta_x+\mu_{xz}
    -i(\eta_y+\mu_{yz})
    \right],
    \qquad
    \beta
    =
    \frac{1}{\sqrt{2}}
    \left[
    \eta_x-\mu_{xz}
    -i(\eta_y-\mu_{yz})
    \right].
\end{equation}
The characteristic equation can be written as
\begin{equation}
    \det(H_{\mathrm{eff}}-\lambda \mathds{1})
    =
    \lambda^3
    -6a^2\lambda
    +i\mu_{xy}(\alpha^*\beta^*-\alpha\beta)
    +\eta_z(|\beta|^2-|\alpha|^2)
    =
    0,
\end{equation}
where we have used that all coefficients have the same modulus \(a\). The corresponding
eigenvector can be written as
\begin{equation}
    \ket{v(\lambda)}
    =
    \frac{1}{N_\lambda}
    \begin{pmatrix}
        -(\eta_z+\lambda)\alpha+i\mu_{xy}\beta^*\\
        \eta_z^2+\mu_{xy}^2-\lambda^2\\
        (\eta_z-\lambda)\beta^*-i\mu_{xy}\alpha
    \end{pmatrix}.
\end{equation}
Using the characteristic equation, one finds that the normalized \(\ket{0}\) component
of these eigenvectors is independent of the eigenvalue \(\lambda\), up to a common
phase convention. Therefore, the \(\ket{0}\) component cancels in the equal-weight
superposition of the eigenvectors with extremal eigenvalues. In particular, one can
choose
\begin{equation}
    \ket{\psi_s}
    =
    \frac{1}{\sqrt{2}}
    \left(
    \ket{v(\lambda_{\max})}
    -
    \ket{v(\lambda_{\min})}
    \right)
    =
    \frac{1}{N}
    \left(
    A\ket{+1}
    +
    B\ket{-1}
    \right),
\end{equation}
which lies entirely in the double-quantum subspace. Hence, the optimal sensing states
for the whole ZENITH family are experimentally accessible by preparing an equal-weight
superposition of the two extremal eigenstates, which reduces to a state in the
\(\{\ket{+1},\ket{-1}\}\) subspace.

\begin{figure}[h]
    \centering
    \includegraphics[width=0.6\linewidth]{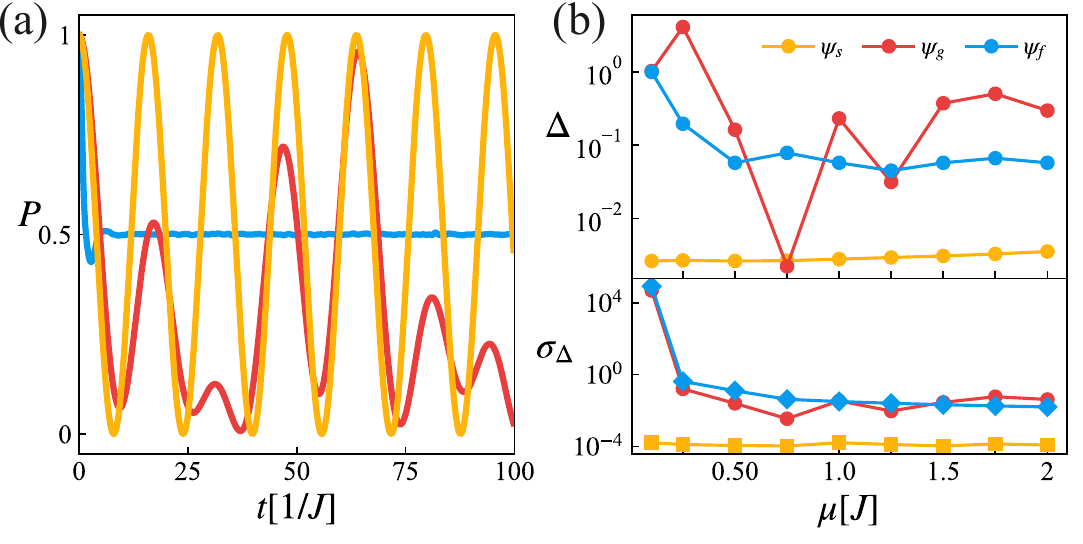}
    \caption{Sensing of a DC signal. (a) Population evolution under a signal coupled to $H_s= 0.5[\text{J}]S_z$. The bright blue line represents the free evolution without the pulse sequence, exhibiting a rapid decay. The red and golden yellow lines represent, respectively, the evolution with the implementation of ZENITH with $\ket{\psi_g}$ and with $\ket{\psi_s}$ as initial states. We can see the lack of decay and the presence of three and one frequency, respectively. (b) Absolute value of the relative deviation from $D$ for different values of such strength, after a least squares fitting (top) and its relative standard deviation (bottom).}
    \label{fig_supp:figure2}
\end{figure}

To demonstrate the usefulness of using the optimal initial state for sensing, we display, in Fig.\,\ref{fig_supp:figure2}, the dynamics for the cases of free induction decay (FID)evolution and ZENITH, where for ZENITH we consider either optimal state preparation ($\ket{\psi_s}$) or a random initial state ($\ket{\psi_g}$). Fig.\,\ref{fig_supp:figure2}(a) shows that both ZENITH initial states oscillate coherently for a long time, whereas FID signal is exponentially suppressed due to a short coherence time. A suboptimal initial state preparation results in a signal that is an admixture of three frequencies oscillating together. The consequences are presented in Figs.\,\ref{fig_supp:figure2}(b) and (c), where a frequency estimation with least squares fitting results in a large error for FID, and unstable estimation for non-optimal state preparation, thereby requiring a much longer evolution time to reach a good estimation. Optimal state preparation yields successful, largely error-free frequency estimation. Importantly, initial state preparation is not resource-intensive as we proved that the optimal state can always be achieved with a single initial pulse.

\par\vskip18pt
\centerline{%
  \rule{0.12\textwidth}{0.5pt}\hskip -1.25em%
  \rule{0.12\textwidth}{1pt}\hskip -1.25em%
  \rule{0.12\textwidth}{1.25pt}\hskip -1.25em%
  \rule{0.12\textwidth}{1pt}\hskip -1.25em%
  \rule{0.12\textwidth}{0.5pt}%
}
\par\vskip18pt
\noindent

\small \noindent\hypertarget{suppref:bertlmann2008}{[S1]} R. A. Bertlmann and P. Krammer,
\href{https://doi.org/10.1088/1751-8113/41/23/235303}{J. Phys. A: Math. Theor. 41, 235303 (2008)}.\\

\small \noindent\hypertarget{suppref:choi2017}{[S2]} J. Choi, S. Choi, G. Kucsko, P. C. Maurer, B. J. Shields, H. Sumiya, S. Onoda, J. Isoya, E. Demler, F. Jelezko, N. Y. Yao, and M. D. Lukin,
\href{https://doi.org/10.1103/PhysRevLett.118.093601}{Phys. Rev. Lett. 118, 093601 (2017)}.\\

\small \noindent\hypertarget{suppref:cerrillo2021}{[S3]} J. Cerrillo, S. Oviedo Casado, and J. Prior,
\href{https://doi.org/10.1103/PhysRevLett.126.220402}{Phys. Rev. Lett. 126, 220402 (2021)}.\\

\small \noindent\hypertarget{suppref:vetter2022}{[S4]} P. J. Vetter, A. Marshall, G. T. Genov, T. F. Weiss, N. Striegler, E. F. Großmann, S. Oviedo Casado, J. Cerrillo, J. Prior, P. Neumann, and F. Jelezko,
\href{https://doi.org/10.1103/PhysRevApplied.17.044028}{Phys. Rev. Applied 17, 044028 (2022)}.\\